\documentclass[twocolumn, times, linenumbers, twocolappendix] {aastex631}

\usepackage{color}
\usepackage{amssymb, amsmath}

\usepackage[utf8]{inputenc}
\usepackage{lipsum}
\usepackage{pifont}
\usepackage{amsmath, amssymb, bm}
\usepackage{balance}
\usepackage{multirow}
\usepackage{mathtools}
\usepackage{enumitem}
\usepackage{booktabs}
\usepackage{appendix}
\usepackage{listings}
\usepackage{graphicx}

\shorttitle{Learning neutrino effects in Cosmology with CNN}
\shortauthors{Elena Giusarma et al.}

\newcommand{\be}{\begin{equation}}
\newcommand{\ee}{\end{equation}}
\newcommand{\ba}{\begin{eqnarray}}
\newcommand{\ea}{\end{eqnarray}}

\newcommand{\mnu}{M_\nu}
\newcommand{\DM}{D$^3$M}

\newcommand{\Quijote}{\textsc{Quijote}}

\hypersetup{citecolor=blue}
\graphicspath{{./}{figures/}}

\begin{document}
\nolinenumbers

\title{Learning neutrino effects in Cosmology with Convolutional Neural Network}

\correspondingauthor{Elena Giusarma}
\email{egiusarm@mtu.edu}


\author[0000-0003-3052-3059]{Elena Giusarma}
\affiliation{Center for Computational Astrophysics, Flatiron Institute, 162 5th Avenue, 10010, New York, NY, USA}
\affiliation{Department of Physics, Michigan Technological University, Houghton, MI 49931, USA}

\author{Mauricio Reyes}
\affiliation{Department of Physics, Michigan Technological University, Houghton, MI 49931, USA}

\author[0000-0002-4816-0455]{Francisco Villaescusa-Navarro}
\affiliation{Center for Computational Astrophysics, Flatiron Institute, 162 5th Avenue, 10010, New York, NY, USA}
\affiliation{Department of Astrophysical Sciences, Princeton University, Peyton Hall, Princeton NJ 08544, USA}

\author{Siyu He}
\affiliation{Center for Computational Astrophysics, Flatiron Institute, 162 5th Avenue, 10010, New York, NY, USA}

\author{Shirley Ho}
\affiliation{Center for Computational Astrophysics, Flatiron Institute, 162 5th Avenue, 10010, New York, NY, USA}
\affiliation{Department of Astrophysical Sciences, Princeton University, Peyton Hall, Princeton NJ 08544, USA}

\author{ChangHoon Hahn}
\affiliation{Lawrence Berkeley National Laboratory, 1 Cyclotron Rd, Berkeley CA 94720, USA}
\affiliation{Berkeley Center for Cosmological Physics, University of California, Berkeley CA 94720, USA}

\begin{abstract}
\nolinenumbers
Measuring the sum of the three active neutrino masses, $\mnu$, is one of the most important challenges in modern cosmology. Massive neutrinos imprint characteristic signatures on several cosmological observables in particular on the large-scale structure of the Universe.  In order to maximize the information that can be retrieved from galaxy surveys, accurate theoretical predictions in the non-linear regime are needed. Currently, one way to achieve those predictions is by running cosmological numerical simulations. Unfortunately, producing those simulations requires high computational resources -- several hundred to thousand core-hours for each neutrino mass case. In this work, we propose a new method, based on a deep learning network (\DM), to quickly generate simulations with massive neutrinos from
standard $\Lambda$CDM simulations without neutrinos. 
We computed multiple relevant statistical measures of deep-learning generated simulations, and conclude that our approach is an accurate alternative to the traditional N-
body techniques. In particular the power spectrum is within $\simeq 6\%$ down to non-linear scales $k=0.7$~\rm h/Mpc.
Finally, our method allows us to generate massive neutrino simulations 10,000 times faster than the traditional methods.

\end{abstract} 


\section{Introduction}
\label{sec:introduction}


The measurement of the mass and the ordering of neutrinos,  
constitutes the necessary steps to understand the physics of those particles 
and the beyond Standard Model mechanism. 
 The discovery of neutrino oscillations~\citep{Fukuda1998,sno2002,kamland2005, minos2008, Ahn2012, abe2012, Abe2014a, Forero2014} have robustly established that at least two out of the three active neutrinos (electron, muon and tau) are massive. However, oscillation experiments only provide bounds on the neutrino mass squared differences. In the minimal neutrino scenario, the best fit values for the solar and atmospheric mass splitting are $\Delta m_{21}^2 \equiv m^2_2 - m^2_1 \simeq 7.6\times 10^{-5}$~$\rm eV^2$ and $|\Delta m_{31}^2| \equiv |m^2_3 - m^2_1|\simeq 2.5\times 10^{-3}$~$\rm eV^2$~\citep{Gonzalez-Garcia2014,Forero2014,Esteban2017}. Since the sign of the atmospheric mass splitting is still unknown, we have two possible orderings of neutrino masses: normal ($\Delta m_{31}^2 > 0$) and inverted ($\Delta m_{31}^2 < 0$). 
 The lower limits on the sum of neutrino masses set by oscillation 
 experiments are~$\mnu \gtrsim0.06$~eV in the normal hierarchy 
 and~$\mnu \gtrsim0.1$~eV in the inverted hierarchy. The study of 
 the endpoint energy of electrons produced in $\beta$-decay places 
 an upper limit on the total neutrino mass at the level 
 of~$\mnu\lesssim 1.1$~eV~\citep{Kraus:2004zw, KATRIN:2019yun}.

Cosmology provides an independent tool to constrain the neutrino masses and ordering. Neutrinos are the second most abundant particles in the Universe, after photons, and leave distinctive signatures on several cosmological observables. They strongly affect the background evolution of the Universe, as well as the evolution of cosmological perturbations. The light massive neutrinos are relativistic in the early Universe and contribute to the radiation energy density. At late time, they became non-relativistic,\footnote{The neutrino non-relativistic transition takes place at $1+z_{\rm nr}\simeq 1890(m_\nu/1)$~eV.} contributing to the total matter density of the Universe. The non-relativistic neutrinos behave as Hot Dark Matter
(HDM), possessing relatively large thermal velocities compared to any other massive particles. They thus cluster only on scales larger than their free-streaming length,\footnote{The free-streaming scale is defined as the scale below which the growth of neutrino perturbations is strongly suppressed. It can be written as:~$k_s \simeq0.018 \sqrt{\Omega_{\rm m}}(\mnu/{1{\rm eV}})^{1/2}$~\rm h/Mpc.} suppressing the growth of perturbations on smaller scales.
The presence of massive neutrinos also affects the anisotropies of the Cosmic Microwave Background (CMB), as these particles may turn non-relativistic around the decoupling period, and gravitational lensing of the CMB. The current upper bound on $\mnu$ is obtained by combining different cosmological data. In particular, the tightest constraints on $\mnu$, which arise by combining CMB temperature and polarization anisotropies measurements with
different observations of the large-scale structure of the
Universe, range from 0.22 eV to 0.12 eV at 95\% confidence level~\citep{palanque-delabrouille_neutrino_2015, Giusarma2016, vagnozzi2017, Zennaro:2017qnp, Giusarma2018, planck2018-1, planck2019}.

Future cosmological surveys, such as Euclid~\cite{EUCLID}, DESI~\citep{Desi}, WFIRST~\citep{WFIRST}, VRO~\citep{LSST}, and CMB-S4~\citep{CMB4}, are expected to improve the constraints on the cosmological parameters. In particular, upcoming galaxy surveys will provide one of the most important sources of information to determine the neutrino masses and their mass ordering. To achieve those goals, accurate theoretical predictions for the spatial distribution of matter and luminous tracers in the presence of massive neutrino will be crucial. A powerful tool to obtain rigorous predictions in cosmology is through cosmological simulations. In particular, N-body simulations are numerical simulations where cosmological matter fluctuations are evolved under only gravity. 
They allow us to make comparisons between observations and theory. They are used to generate mock galaxy catalogs and to compute error covariance matrices. Lastly, they are also used to optimize observational strategies. Over the past years, a large number of N-body simulations, including massive neutrinos, have been proposed and developed in the literature. Those simulations have made it possible to study the impact of neutrino masses on clustering in fully non-linear scale in real-space~\citep{Bird:2011rb, Lesgourgues:2012uu, VillaescusaNavarro:2011me, Paco2014, Peloso:2015jua}, on clustering and abundance of halo and cosmic voids~\citep{Castorina2014, Castorina:2015bma, Massara_2015} and finally on clustering of matter in redshift-space~\citep{Villaescusa-Navarro:2017mfx, Bel:2018awq}. However, simulations are computationally expensive. The development
of new computational methods is thus needed to accelerate this process. 

In the last decade, deep learning approaches have been applied to many different fields of research, achieving outstanding results; especially in computer vision and image recognition challenges~\citep{imagenet_2012}. Deep learning is a branch of machine learning that makes use of deep neural networks. An increasing number of studies are also adopting these methods for a variety of problems in cosmology. For example,~\cite{ravanbakhsh_estimating_2017},~\cite{he2018analysis} and ~\cite{Ntampaka:2019ole} applied convolutional neural networks (CNNs) to estimate the cosmological parameters directly from simulated dark matter distributions and the distribution of photons in Cosmic Microwave Background;~\cite{He2019} built a deep CNN to predict the non-linear structure formation of the Universe from simple linear perturbation theory;~\cite{Rodriguez:2018mjb} and~\cite{Zamudio-Fernandez:2019lxp} used Generative Adversarial Networks (GANs) models to synthesize samples of the cosmic web and predict high-resolution 3D distributions of cosmic neutral hydrogen;~\cite{Ramanah:2019cbm} introduced a novel halo painting network to map the cosmic density fields of dark matter to realistic halo distributions;~\cite{Hezaveh:2017sht} exploited the use of CNNs to estimate the strong gravitational lensing parameters accurately; finally,~\cite{List:2019msk} applied conditional Generative Adversarial Networks (cGANs) to predict how changing the underlying physics alters the simulation results.

In this paper, we apply a modified version of the deep neural network proposed by~\cite{He2019}  to establish the mapping between simulations with massless and massive neutrinos. The important advantage of our approach consists in the ability to produce complex numerical simulations from the standard ones in a fraction of few minutes using a modern Graphics Processing Unit (GPU), and with the same statistical properties as the standard N-body techniques. 

The paper is organized as follows. In Section~\ref{sec:Simulations} we briefly describe the cosmological simulations we have used in this work. We then present a description of the training process and the quantitative results in Sections~\ref{sec:Model1} and~\ref{sec:Results}. Finally, we draw our conclusions in Section~\ref{sec:Conclusions}.

\section{N-body simulation data}
\label{sec:Simulations}
In this section, we briefly describe the N-body simulations used in our work. We refer the reader to \cite{Villaescusa-Navarro:2017mfx, Villaescusa-Navarro19} for further details on the simulations.

We use two subsets of HADES simulations \citep{Villaescusa-Navarro:2017mfx}, the precursor of the \textsc{Quijote} simulations \citep{Villaescusa-Navarro19}: the standard, $\Lambda$CDM, simulations (without neutrinos), and simulations with massive neutrinos. The value of the cosmological parameters are: $\Omega_{\rm m}=\Omega_{\rm cdm}+\Omega_{\rm b}+\Omega_\nu=0.3175$, $\Omega_b = 0.049$, $h=0.6711$, $\Omega_\Lambda = 0.6825$, $n_s=0.9624$ and $A_s = 2.13$. The simulations with massive neutrinos have $M_\nu=0.15$ eV, and since $A_s$ is fixed in both sets, the value of $\sigma_8$ is different: $\sigma_8=0.833$, and $\sigma_8=0.798$ for massless and massive neutrinos respectively. All the simulations are run in a periodic box size of $1~h^{-1}{\rm Gpc}$ and follow the evolution of the cold dark matter (CDM) and neutrino particles (only in the case of simulations with massive neutrinos), from $z=99$ to $z=0$.  We use 100 independent realizations for each model, containing the same number of cold dark matter and neutrino particles ($N_{\rm cdm} = N_\nu =512^3$). We focus our analysis at redshift $z=0$, corresponding to the present time. 

In order to speed up the process of training the neural network and to avoid GPUs memory problems, we split each of the 100 simulations into sub-cubes of size $32^3$ voxels in Lagrangian space,\footnote{Notice that the initial conditions of our simulations are generated from a regular grid.} corresponding to~regions of size $62.5~ h^{-1}{\rm Mpc}$. Each realization contains thus $4,096$ sub-cubes, making a total of $409,600$ voxels for each simulation model. We then split the realizations of each model into three chunks: $70\%$ for training (70 realizations), $20\%$ for validation (20 realizations) and $10\%$ (10 realizations) for testing. After the training and the testing, we concatenate the sub-cubes of each of the 10 realizations in the test data in a cube of size $1~h^{-1}{\rm Gpc}$ and we compute the relevant summary statistics. 

\section{Model}
\label{sec:Model1}
\subsection{Model and Training} 
\label{sec:Training}
We use a modified version of the deep neural network, Deep Density Displacement Model (D$^3$M), introduced by~\cite{He2019} with 15 convolution plus deconvolution layer. D$^3$M is a generalization of the standard U-Net, first proposed in~\cite{U-Net} for biomedical image segmentation, to work with three-dimensional data. A brief description of the architecture is shown in Appendix~\ref{sec:appendixA}.

Following~\cite{He2019}, we train our model using the displacement field rather than density field (see~\cite{He2019} for more details). The displacement field is  defined as:
\be
\vec{d}=\vec{x}_f-\vec{x}_i~,
\ee
where $\vec{x}_f$ is the final position of the particles (at redshift $z=0$) and $\vec{x}_i$ is the Lagrangian position of the same particles on a uniform grid. The input and target~\footnote{In machine learning the target is defined as the desired response we are trying to predict. In particular, the training is performed by evaluating the model using a data sample for which we already know the result.} of the neural network is the displacement field from simulations without and with massive neutrinos, respectively. 

Our benchmark model consists of the Zel'dovich approximation (hereafter ZA) applied to massive neutrino models. The ZA is a significantly faster approximation to full N-body simulations produced by first-order perturbation theory with neutrinos. The reason why we use the Zel’dovich approximation and not the second-order Lagrangian perturbation theory (2LPT), is because it is currently unknown how to predict 2LPT in massive neutrino models. In particular, there is no estimate for the two significant quantities, the second-order scale-dependent growth factor, and growth rate, necessary to calculate the 2LPT in presence of massive neutrinos.   

We train our neural network by using the mean absolute error loss function that is given by the sum of all the absolute differences between the target displacement field and the predicted values: 
\be
\mathbb{L} = {\sum_{i=1}^N \left\vert\vec{d}_i^t-\vec{d}_i^p\right\vert}~,
\ee
where $\vec{d}_i^t$ is the true displacement field (from massive neutrino simulations), $\vec{d}_i^p$ is the prediction from the \DM~, and $N$ is the total number of particles. 

The total number of trainable parameters is $8.4\times10^6$.


\begin{figure*}[t]
\begin{center}
\textbf{\hspace{0.5cm}\large{Target}\hspace{4.5cm}\large{Target $-$ Input~}\hspace{3cm}\large{Target $-$ \DM~}}\par\smallskip
\includegraphics[width=.33\linewidth]{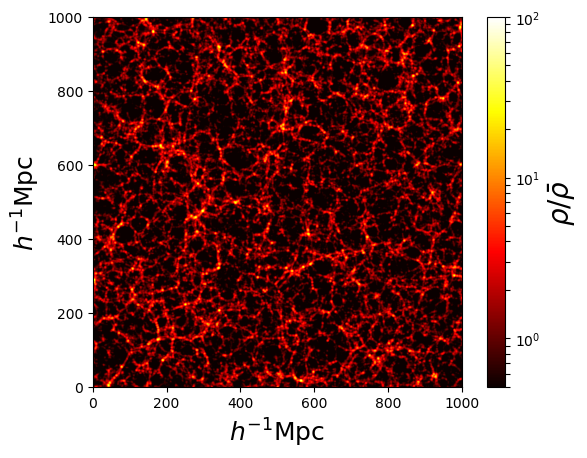}
\includegraphics[width=.33\linewidth]{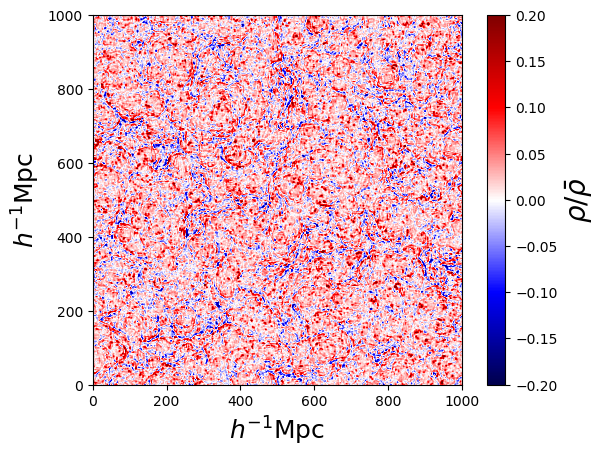}
\includegraphics[width=.33\linewidth]{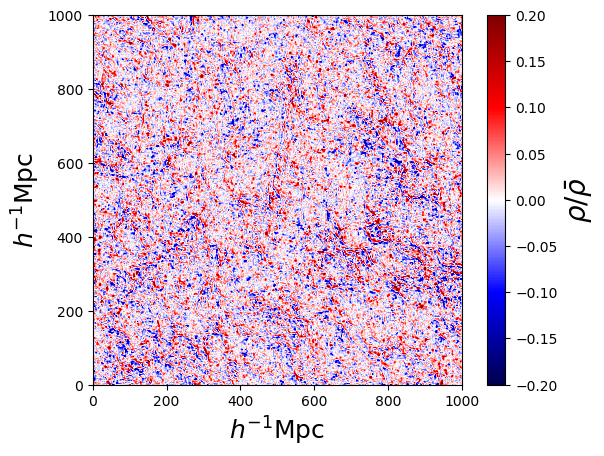}\\
\vspace{0.5cm}
\includegraphics[width=.33\linewidth]{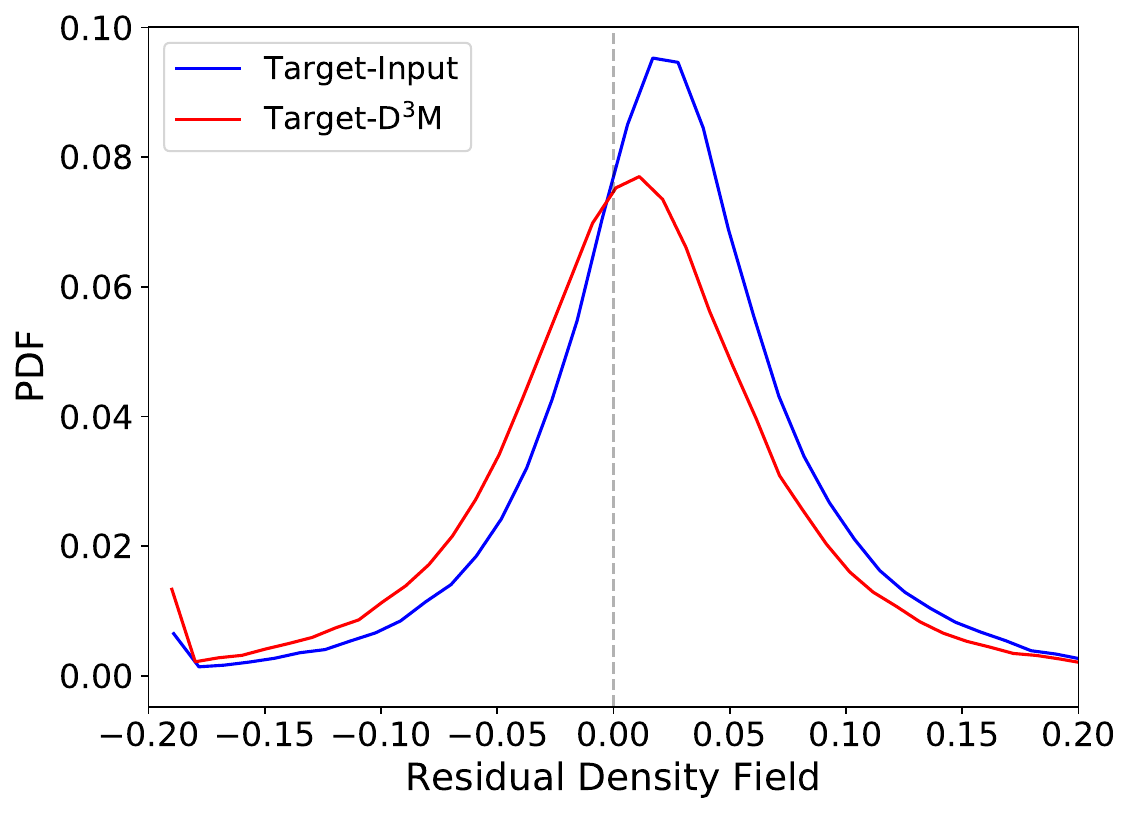}
\hspace{3cm}
\includegraphics[width=.33\linewidth]{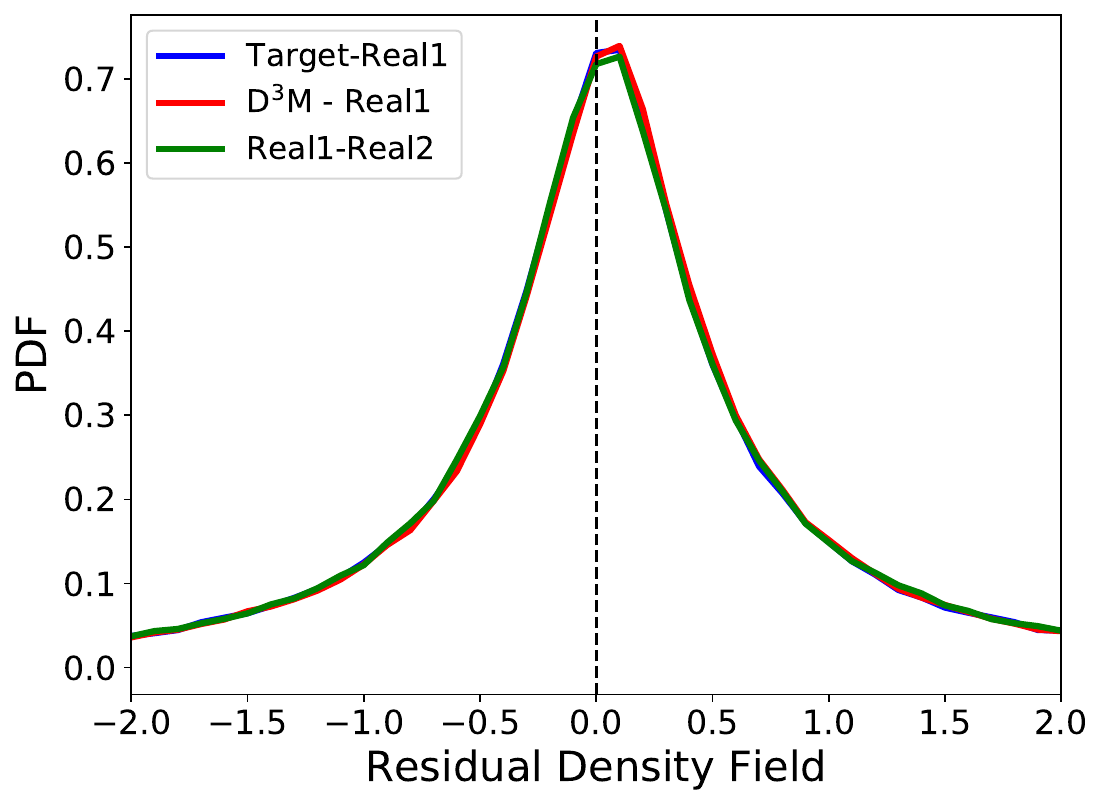}\\
\caption{The top left panel shows the cold dark matter density field at $z=0$ from a region of $1000\times1000\times15$ (h/Mpc)$^3$ for a massive neutrinos model ($\mnu=0.15$~eV). The top-center panel shows the residuals between the target (massive neutrino simulations) and the simulations with massless neutrinos (the input); the top-right panel depicts the residual between the target and \DM. The two bottom panels illustrate the PDF of the residual density field as described in the text.}
\label{fig:density}
\end{center}
\end{figure*}

Figure~\ref{fig:density} shows the projected density field of  N-body simulations with massive neutrinos at $z=0$ (top-left panel), together with the residuals between the target and the input (top-center panel) and the target and the output of our neural network (top-right panel). From the top-left panel, we can clearly see the presence of the filaments, surrounded by large underdense regions (dark regions), called voids, in between. The presence of massive neutrinos in the simulations induces a suppression in the growth of density perturbations on small scales, visible in the matter power spectrum. 

The bottom-left panel of Figure~\ref{fig:density} shows the PDF of the residuals from ``Target $-$ Input" (blue line) and ``Target $-$ \DM" (red line). By computing a Q-Q plot test, we established that both the curves approximately follow a Gaussian distribution. However, the blue curve exhibits a shift induced by the presence and the absence of neutrinos in the simulations. In order to verify if the statistical variations between the simulations is reflected in the
\DM~prediction, we also consider a different simulation instance in the training data with $\mnu=0.15$~eV, see bottom-right panel of Figure~\ref{fig:density}. In particular, by comparing  the residuals between ``\DM $-$ realization 1'' (red line), ``Target $-$ realization 1'' (blue line) and ``realization 1 $-$ realization 2'' (green line),  we can note that the three residual distributions present a same shape and width. This is an indication that the density residuals coming from our \DM~model have similar statistical properties as the simulations with the same neutrino mass. 

\subsection{Training time} 
\label{sec:Training time}

One standard cosmological simulation without neutrinos currently requires $\sim500$ 
CPU hours to complete. 
The addition of massive neutrinos increases the simulation time to $\sim700$ 
CPU hours. 
The key advantage of the approach we describe in this work is its ability to reduce 
the additional computational cost of producing massive neutrino simulations.
In particular, it requires $\sim10$ GPU hours to train and only few minutes 
to generate one simulation with massive neutrinos: a {\em substantial} gain in
computing time compared to standard N-body techniques. We run our CNN model on two GPUs,~320 NVIDIA P100-16GB. This allows us to speed up the training time by a factor of 2.5.


\section{Results}
\label{sec:Results}
In this section, we present the main results obtained by using the \DM~ architecture and compare it with the standard N-body simulations and the benchmark model. We quantify the agreement between our model and numerical simulations by considering four different summary statistics: the overdensity distribution function, the power spectrum, the bispectrum, and the void size function. By doing this, we can compare the properties of the fully 3-dimensional density distributions.
We compute the considered statistics by concatenating the sub-cubes of each of the ten realizations in the test data and calculating the mean summary statistics over ten simulations.

\subsection{1-D Probability distribution function}

We compute the overdensity distribution of the cold dark matter for the N-body simulations and \DM~prediction. We first place the particles on a grid with $512^3$ cells, and we smooth the density field with a top-hat filter on a scale of 10 Mpc/h. We then calculate the distribution as the fraction of cells with overdensity lying within a given interval. The probability distribution function provides information on the distribution of CDM overdensities in the cells. 

\begin{figure}[t]
\vspace{0.3cm}
\centering
\includegraphics[width=0.47\textwidth]{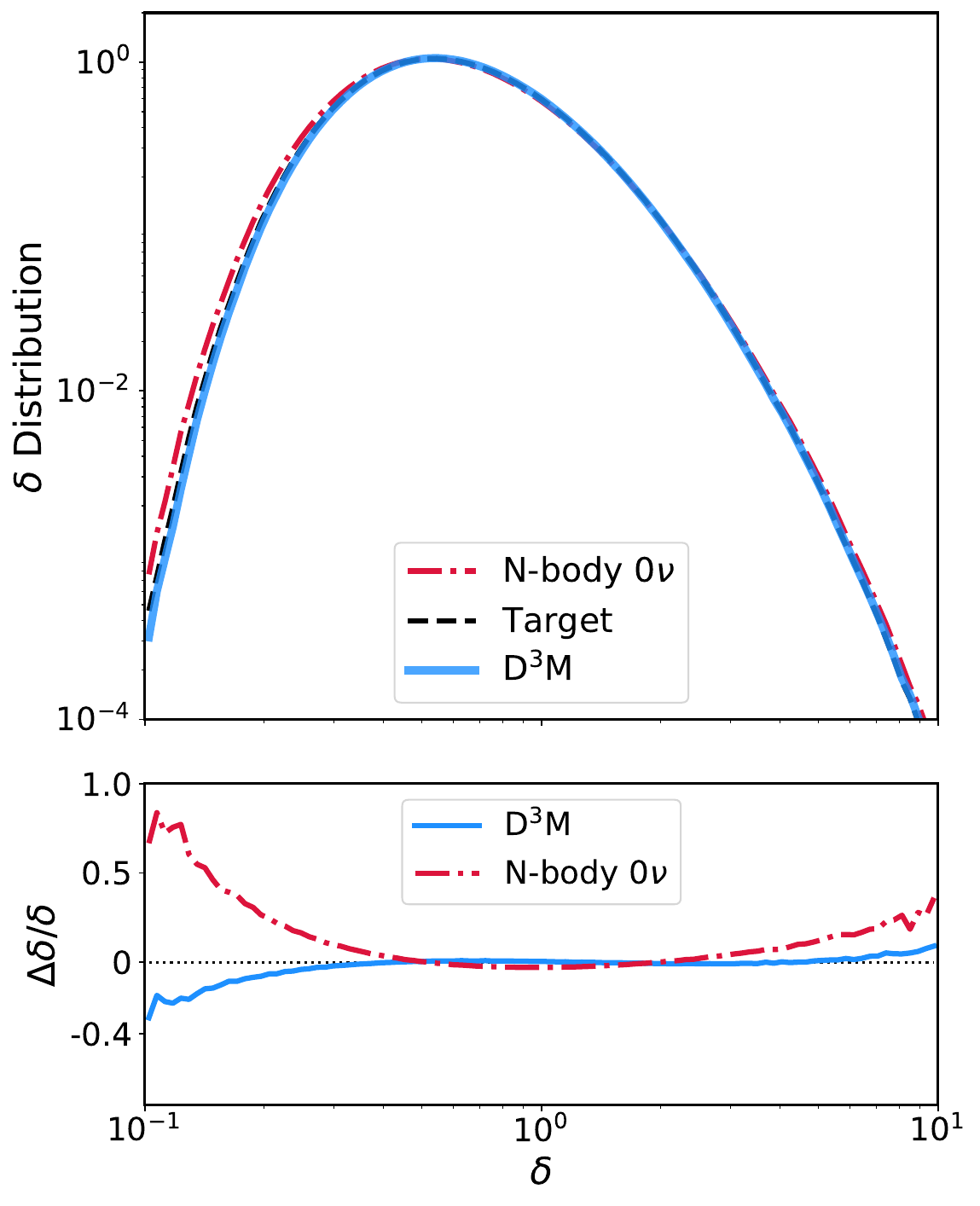}  
\caption{The top panel shows the normalized probability distribution function of CDM overdensities. The red and black lines show the overdensity distribution for N-body simulations with massless (the input of our \DM) and massive neutrinos (target), respectively. The blue line depicts the \DM~results. The bottom panel illustrates the residual for the \DM~with respect to the simulations with massive neutrinos.}
\label{fig:PDF}
\end{figure}

The results are depicted in Figure~\ref{fig:PDF}, where we show (upper panel) the comparison of the overdensity distribution among the standard simulations (the input, red dashed-dotted line), massive neutrino simulations (the target, black dashed line) and our \DM~model (blue line).\footnote{Here the overdensity is defined as the ratio of the density field smoothed over a scale of 10 Mpc and the global average
density for the simulation.}
The bottom panel of Figure~\ref{fig:PDF} shows the residual of the neural network with respect to the target. The residual is approximately 0 for $0.3<\delta<6$ and differs from zero by a $15\%$ and $6\%$ for $\delta \approx 0.15$ and $9$ respectively. Those results indicate a good agreement between the \DM~and the target.

\subsection{Power Spectrum}
\label{sec:Power Spectrum}
The most widely used statistic in cosmology to extract information from cosmological observations is the 2-point correlation function $\xi(r)$, which is defined as the excess probability, compared with a random distribution of galaxies, of
finding a pair of galaxies at a given separation. The Fourier transform of the 2-point correlation function is the power spectrum, $P(k)$: 

\be
\label{eq:1}
\begin{aligned}
\xi(r)= \langle \delta_{\rm}(\vec{r}') \delta(\vec{r}'+\vec{r})\rangle 
\\
P(k) = \int d^3 \vec{r} \xi(r) e^{i\vec{k}\cdot\vec{r}}~,\\
\end{aligned}
\ee
where $k$ is wavenumber of a fluctuation, that is related to the wavelength $\lambda$ by $k = 2\pi/\lambda$, and $\delta(\vec{r})$ is the overdensity at position $\vec{r}$ defined as $\delta=\rho/\bar{\rho}-1$ with $\bar{\rho}$ denoting the mean density.~\footnote{In equation~\ref{eq:1}, we make the assumption that the density field is statistically homogeneous and isotropic. This implies that the correlation function, $\xi(\vec{r})=\xi(r)$, is a function only
of the scalar separation $r = |\vec{r}-\vec{r}'|$ of the points $\vec{r}$ and $\vec{r}'$, not of their overall location or orientation.} The power spectrum is the quantity predicted directly by theories for the formation of large-scale structure, and in the
case of a Gaussian density field, it provides a complete statistical description of its fluctuations.

In order to quantify the performance of our model against the target, we compute the transfer function $T(k)$, defined as the square root of the ratio of the two power spectra: 
\be
\label{eq:2}
T(k)=\sqrt{\frac{P_{\rm pred}(k)}{P_{\rm target}(k)}}~, \\
\ee
where $P_{\rm pred}(k)$ and $P_{\rm target}(k)$ are the power spectra of the density field predicted by the \DM~~ and the target, respectively.

\begin{figure}[t]
\vspace{0.25cm}
\begin{center}
\includegraphics[width=0.5\textwidth]{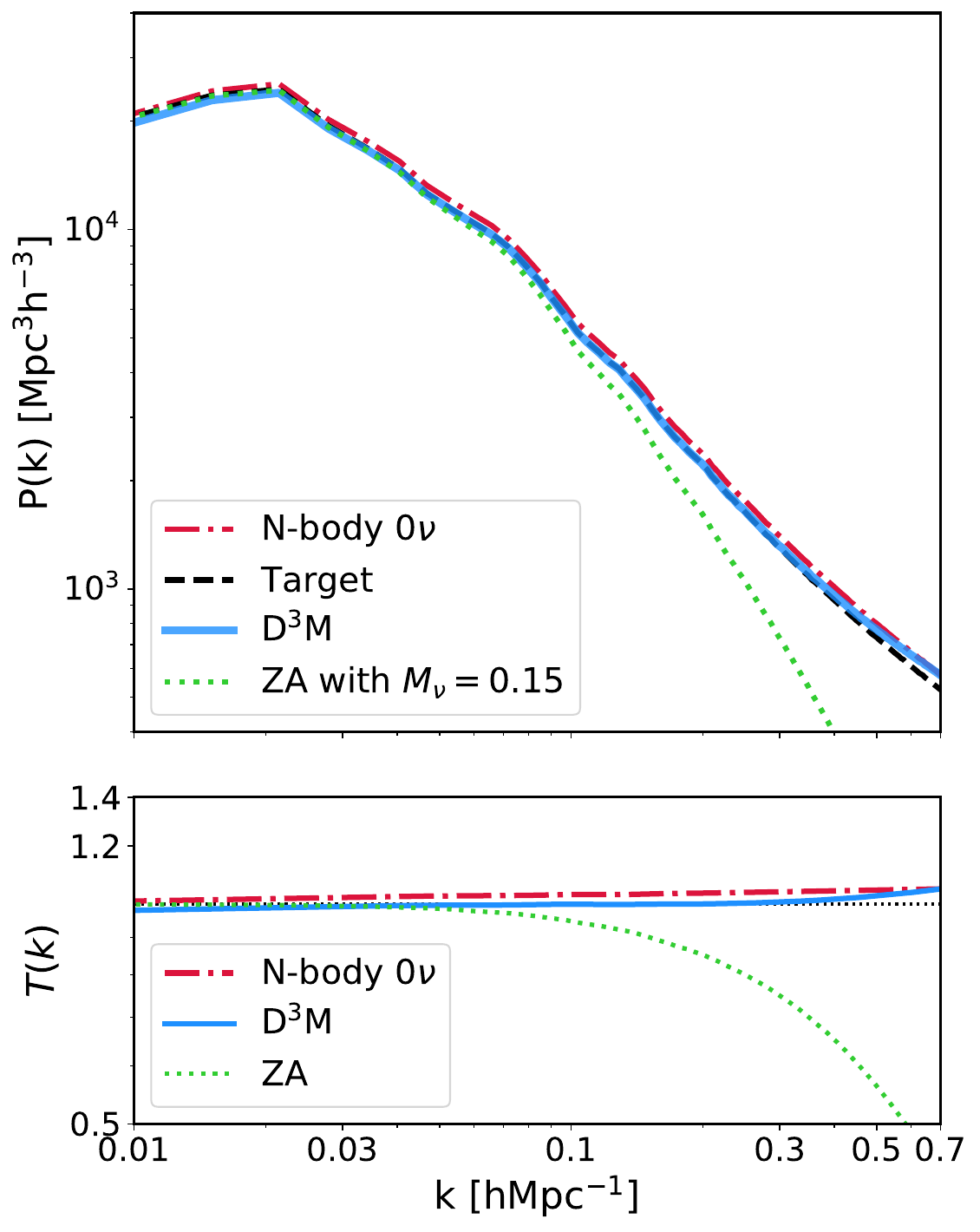}  
\caption{The top panel shows the matter power spectrum comparison among the standard N-body simulations (the input of the \DM, red dashed-dotted line), massive neutrino simulations (the target, black dashed line), CNN model dubbed \DM~(blue line) and Zel'dovich approximation (the benchmark model, green dotted line). The bottom panel depicts the transfer function for the input, the \DM~and the benchmark model.}
\label{fig:Pk}
\end{center}
\end{figure}

Figure~\ref{fig:Pk} shows the average power spectrum and the transfer function of the density field over ten simulations.  The red and black lines on the top panel correspond to the power spectrum of N-body simulations without and with massive neutrinos. We note that massive neutrinos induce a suppression of the power spectrum at small scales (larger $k$). This is the most significant effect of neutrino masses on the several cosmological observables. Neutrinos, being hot thermal relics, possess large velocity dispersion. Consequently, the non-relativistic neutrino overdensities will only cluster at wavelengths larger than their free streaming wavenumber, reducing the growth of cold dark matter fluctuations on small scales. 

The blue and green lines depict the power spectrum from the \DM~ and benchmark model (Zel'dovich approximation). 
Notice that the latter reproduces the power spectrum accurately on very large-scales 
($k<0.05$~$h$/Mpc). This is what we expect since we are using the Zel’dovich approximation 
and not 2LPT (see subsection~\ref{sec:Training}). 

We find that the density distribution from the \DM~ sample has, on large-scales an average power spectrum which is very close in amplitude and in shape to that of N-body simulations with neutrinos (the target). 
On smaller scales, the power spectrum of our model departs from that of the target. To quantify this deviation, we focus on the bottom panel of Figure~\ref{fig:Pk}, that shows the transfer function as a function of wavenumber for the benchmark model, the \DM~and the input. For an entirely accurate prediction, the transfer function is expected to be 1. The transfer function of the ZA approximation is close to one up to $k<0.05$~h/Mpc and then decreases as expected. On the other hand, the transfer function of the \DM~ prediction is approximately 1 for $0.03<k<0.5$~h/Mpc and differ from the unity by a $3.5\%$ on scales $k\approx0.55$~h/Mpc. This discrepancy increases to $6.3\%$ as $k$ increases around 0.7~h/Mpc. Those results suggest that our model manages to predict the power spectrum accurately with massive neutrinos from large to intermediate scales. On smaller scales ($k>0.7$~h/Mpc) the prediction starts to deviate from the target. This is not a surprise since those are the scales where the effects of non-linear evolution are important, and the mapping becomes highly non-trivial.

\subsection{Bispectrum}
\label{sec:Bispectrum}
\begin{figure}[t]
\vspace{0.3cm}
\centering
\includegraphics[width=0.47\textwidth]{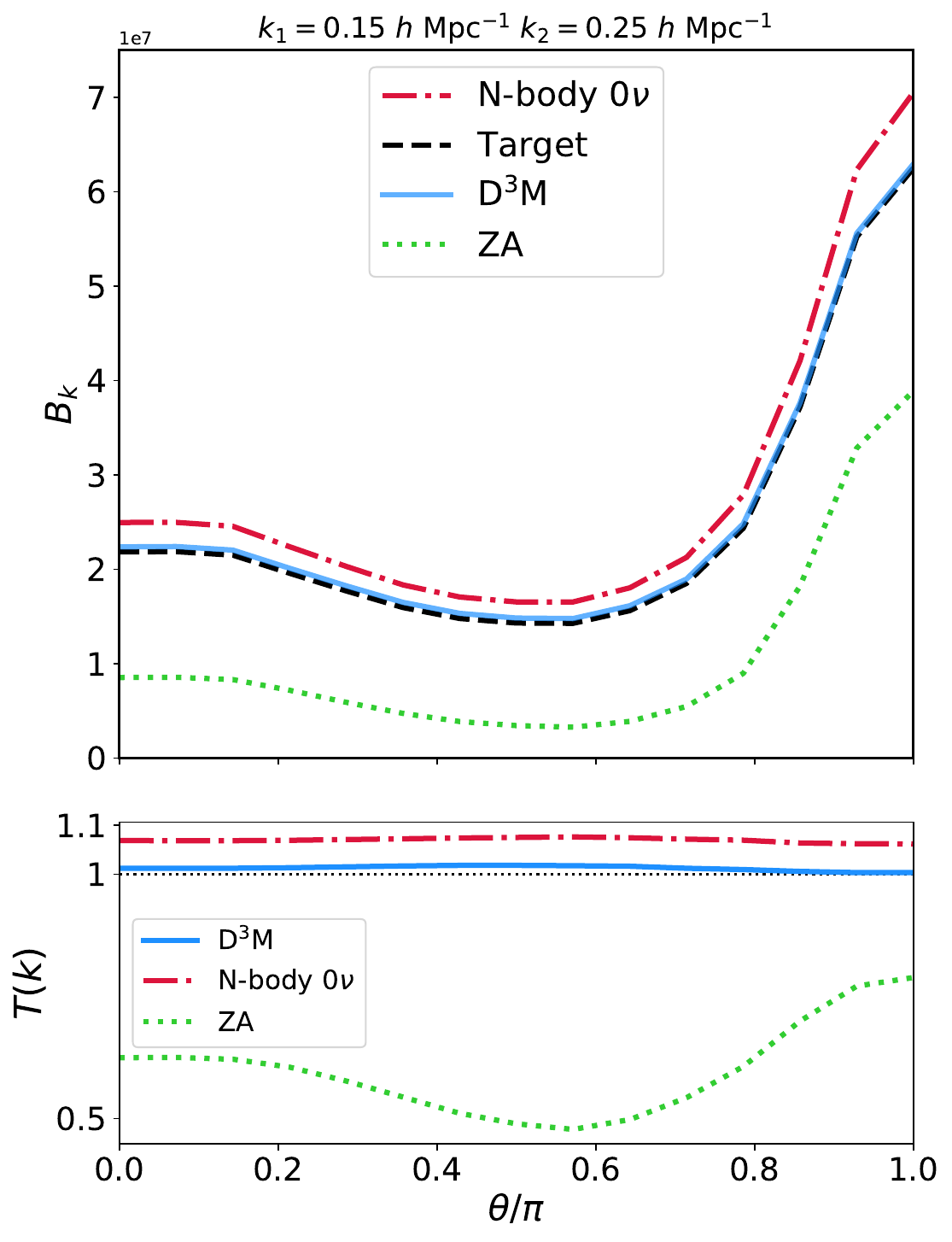}  
\caption{The plot shows the bispectrum predicted by standard $\Lambda$CDM simulations (the input of our architecture, red dashed-dotted line), massive neutrino simulations (the target, black dashed line), CNN model (blue line) and Zel'dovich approximation (the benchmark model, green dotted line). We set $k_1=0.15$~h/Mpc and $k_2=0.25$ h/Mpc. Notice that the bispectrum of the \DM~ model is very close to that of the target.}
\label{fig:Bk}
\vspace{0.3cm}
\end{figure}

The inflationary scenario predicts Gaussian initial conditions. The statistical properties of the Universe's density field should be fully characterized by the 2-point correlation function (2PCF) or the power spectrum. However, on small scales, non-linear gravitational instability induces non-Gaussian signatures in the mass distribution, which contain information on the nature of gravity and the dark matter. The lowest order statistical tool to describe a non-Gaussian field is the three-point correlation function (3PCF), or
equivalently, its Fourier transform, the bispectrum, defined as:

\be
B(k1, k2, k3)=\langle\delta_{\bf k_1}\delta_{\bf k_2}\delta_{\bf k_3}\rangle~.
\ee

The bispectrum plays an important role in cosmology because it carries information about the spatial coherence of large-scale
structures and can place strong constraints on models of structure formation. The presence of massive neutrinos induces a suppression of the amplitude of the bispectrum~\citep{Ruggeri:2017dda}. 
Furthermore, \cite{Hahn:2019zob} have shown that the bispectrum is a powerful probe in 
constraining the total neutrino mass since can break the degeneracies between $\mnu$
and the other cosmological parameters. 

Figure~\ref{fig:Bk} shows the average bispectrum over ten simulations for our \DM~ (blue line), the benchmark model (green line), the input (red line) and the target (black line). We have selected a value of $k_1=0.15$~h/Mpc, and of $k_2=0.25$ h/Mpc and varied the angle among those two vectors. Those scales correspond to the maximum wavenumbers adopted in cosmology to infer the cosmological parameters from the galaxy power spectrum measurements. 

We find that the mean relative bispectrum residual of our model compared to the target is $0.4\%$. This suggests that the \DM~ model can generate neutrino simulations with correct higher-order statistics in the mildly non-linear regime. Notice that on the other hand, the benchmark model is not able to reproduce the bispectrum from the target. This is not surprising as the validity of the Zel'dovich approximation, will be limited to large scales.

\subsection{Voids Abundance}
\label{sec:Voids}
Voids are the most under-dense regions of our Universe and together with halos, filaments, and walls, constitute the large-scale structure of the Universe, known as the cosmic web. Voids enclose a large amount of cosmological information since they fill most of the volume of the Universe. Massive neutrinos modify the properties of cosmic voids, such as their number, size and shape. This occurs because the large thermal velocities of massive neutrinos prevent their evacuation from cosmic voids. Consequently, such additional mass within voids affects their evolution. 

\begin{figure}[t]
\vspace{0.3cm}
\centering
\includegraphics[width=0.48\textwidth]{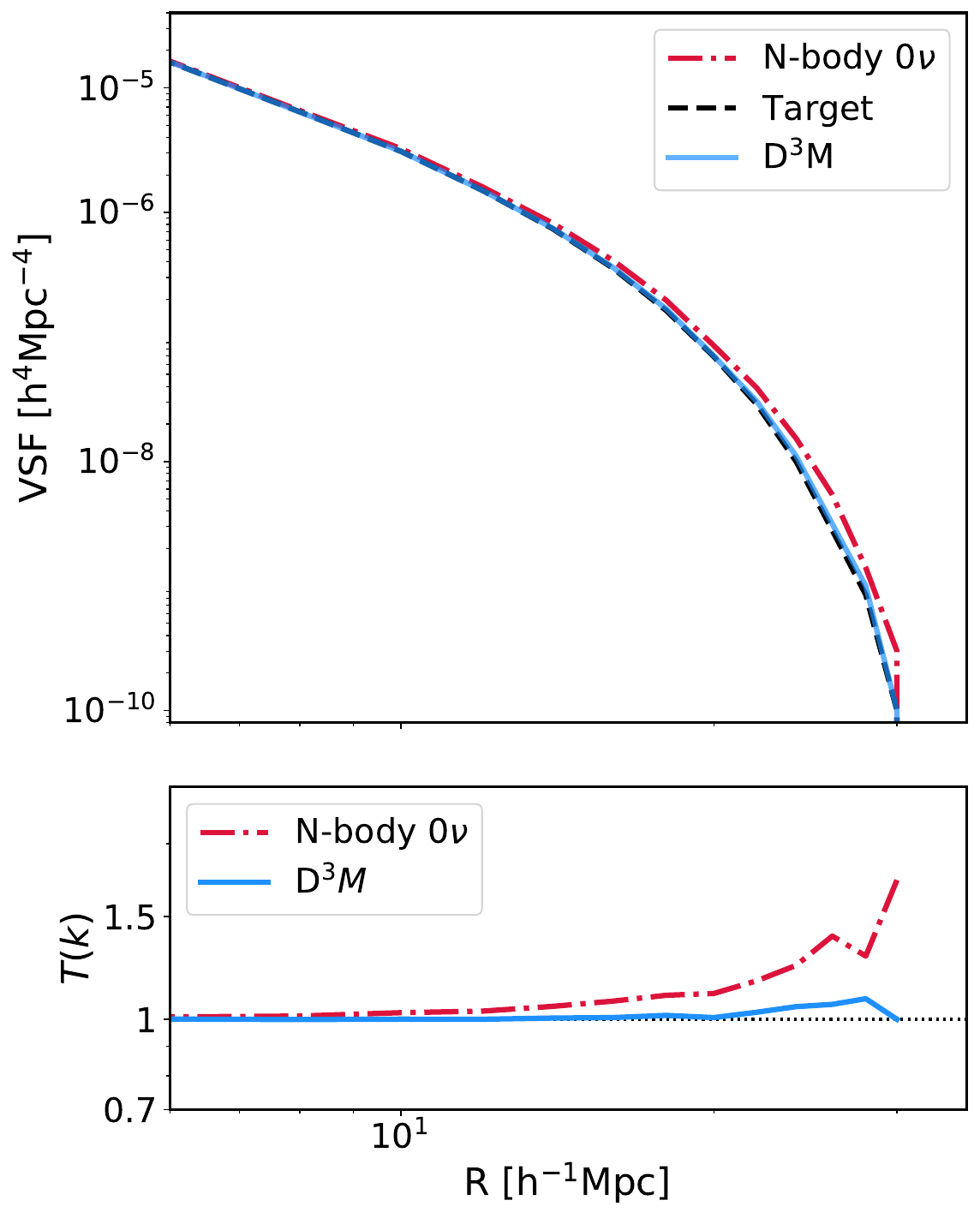}  
\caption{The upper panel shows the void size function for the input (standard N-body simulations, red dashed-dotted line), target (massive neutrino simulations, black dashed line) and \DM~(blue line). The bottom panel depicts the transfer function for the input and the neural network model.}
\label{fig:VF}
\vspace{0.3cm}
\end{figure}

The last statistical tool we consider in this section is the void size function, defined as the abundance of voids as a function of the voids radius. In presence of massive neutrinos, the abundance of large voids is suppressed. This occurs because the additional neutrino masses inside the voids slow down their evolution and make voids smaller and denser~\citep{Massara_2015, Kreisch:2018var, Pisani:2019cvo}. Here we used the algorithm described in~\cite{Banerjee_2016} in order to identify voids in the galaxy distribution of our \DM, the target and the input. In Figure~\ref{fig:VF}, upper panel, we compare the average void size function over ten realizations from the \DM~(blue line), the input (red line), and the target (black line). We can see the suppression of the voids abundance induced by massive neutrinos (blue and black lines) at larger radius. The bottom panel shows the transfer function as a function of radius for the input and the neural network model. We can note the good agreement in the void size function among the \DM~and the target up to $R\approx20$ $[\rm h^{-1} Mpc]$. The discrepancy from the unity increases by a $8\%$ as the radius increases around 30 $[\rm h^{-1} Mpc]$. This indicates that our approach is also able to capture voids' information at the relevant scales.

\subsection{Additional tests}
\label{sec:Addtests}
Although our model reproduces the target's power spectrum within $6.3\%$ up to scales $k\approx0.7$~h/Mpc, the accuracy degrades on smaller non-linear scales. In order to improve the model efficiency at those scales, we perform some additional tests. 

We first tune the hyperparameters in the neural network. In particular, we increase the number of the layers, change the learning rate and the batch size, and apply the mean square error (MSE) loss function. 

We then test different convolutional neural network architectures. Specifically, we implement the \DM~ by including an Inception module after the first block in the down-sampling and before the last block in the expansive path.
The Inception network~\citep{Inception} uses convolutions of different sizes to capture details at varied scales. In our case, we are interested in extracting detailed features on smaller scales. This motivates the use of the Inception module at the beginning and at the end of the \DM~. In particular, we use the first original version of the Inception architecture with 3 different sizes of filters ($1\times1$, $3\times3$, $5\times5$) and max pooling layers. The outputs are then concatenated and passed as input to the second/last block of the \DM~.

\begin{figure}[t]
\centering
\includegraphics[width=0.47\textwidth]{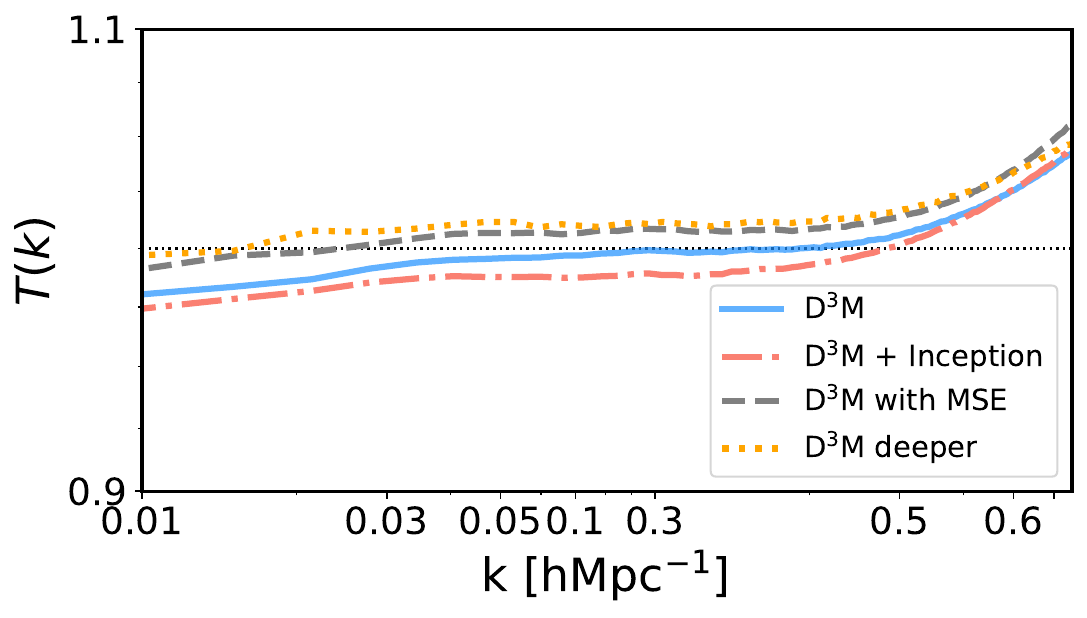}  
\caption{Transfer function comparison among the \DM~ (blue line) used in the analysis and the modified version of the architecture obtained by including the Inception module (pink dashed-dotted line), varying the loss function (grey dashed line) and increasing the number of layers of about the double of \DM~(orange dotted line).}  
\label{fig:Transfer}
\vspace{0.3cm}
\end{figure}  
The results are depicted in Figure~\ref{fig:Transfer}. We compare the transfer function of the additional tests we performed with the \DM~ architecture used in our analysis. 
We can see that these tests do not significantly improve the performance 
of our model at smaller scales. We only find a slight improvement on larger 
scales when we consider the MSE loss function or a deeper \DM~ architecture. Moreover the agreement for small scales is compromised in favor of the agreement for large scales when we use a deeper \DM~model. This demonstrates the stability of our model and shows that more accurate 
trials are needed to improve the predictions on non-linear scales.
We leave this exploration to future work. 
It is lastly worth to mention that in order to speed up the learning process of our \DM~ architecture, we parallelized the training across the data dimension using two GPUs. In particular, to train on multiple GPUs we used the PyTorch multi GPU techniques—data parallelism. By doing that, the two GPUs synchronize the model parameters (the weights) to ensure that they are training a consistent model. Unfortunately, such data parallelism suffers from excessive inter-GPU communication overheads due to frequent weight synchronization among GPUs. This process introduces some systematic during the training that affects the final prediction (see for example the orange dotted line in Figure~\ref{fig:Transfer}). The development of new  parallelization approaches to speed up the training process in deep neural networks and improve the predictions is one of major research area of study in machine learning~\citep{Chen2018EfficientAR, Krizhevsky2014OneWT, lin2018deep} and it is well beyond the scope of our paper.\

\section{Conclusions}
\label{sec:Conclusions}
Upcoming galaxy surveys, such as Euclid, DESI, DES or LSST, are expected to reach the sensitivity necessary to detect the absolute scale of neutrino masses. To achieve this goal, precise theory predictions are required to analyze the data from those surveys and to understand and model the impact of non-linearities in the matter power spectrum, galaxy bias, and redshift space distortion in presence of massive neutrinos. A powerful way to obtain accurate theory predictions is by running numerical cosmological simulations. However, producing those simulations requires high computational resources. 

In this work, we have presented a deep learning approach to predict fast non-standard cosmological simulations with massive neutrinos from standard N-body simulations. We have quantified the performance of our \DM~ model by considering four summary statistics often used in cosmology: the 1-D PDF, the power spectrum, the bispectrum, and the void abundance. We have shown that our model is able to learn the mapping between the displacement field in simulations with massless and massive neutrinos, down to scales as small as 0.7 $h$/Mpc at $z=0$, and reproduce the effect of massive neutrinos in large-scale structure.
Moreover, with our method, we can generate massive neutrino simulations 
five orders of magnitude faster than the traditional methods and dramatically 
reduce the computational costs necessary of implementing simulations with 
massive neutrinos.

This work demonstrates that the deep learning approach is an accurate alternative 
to the traditional N-body techniques by mapping simulations with massive neutrinos 
directly from simulations with massless neutrinos. The use of such approach will 
be particularly useful in the near future. The need for fast N-body simulations 
will increase with the upcoming large cosmological surveys such as Euclid and LSST. 
Our methods will thus be crucial for producing fast and large simulation datasets 
for cosmological analyses.

In future work, we will improve our model to capture the effect of massive 
neutrinos on smaller non-linear scales by using the $\Quijote$ high-resolution simulations~\citep{Villaescusa-Navarro19}. Another direction will be to 
explore the possibility of generating fast neutrino simulations with an arbitrary 
mass from the standard $\Lambda$CDM simulations or more complex non-standard 
cosmological simulations that include, for example, modified gravity or primordial 
non-gaussianity effects~\citep{Peel:2018aei, Novaes:2014ska}. We leave
for future work the investigation of these different effects. 
We note that recently, \cite{Angulo} has presented a new method to \textit{rescale} massless neutrinos simulations to massive neutrinos simulations, following a very different approach to the one used in this work. Using the output of that method as the input of our neural net has the potential to improve the accuracy of both our method and \cite{Angulo} results, as neural nets train faster on residuals. We plan to study this in future work.

\section*{ACKNOWLEDGEMENTS} 

We thank Gabriella Contardo, Elena Massara, Barnabas P\'ocsoz, Siamak Ravanbakhsh, and David Spergel for useful discussions. EG acknowledges Nick Carriero from the Flatiron Institute for his help in solving technical problems with GPUs. The work of EG, FVN, SH and SH is supported by the Simons Foundation. The \textsc{HADES} and \textsc{Quijote} simulations are publicly available at \url{https://franciscovillaescusa.github.io/hades.html} and \url{https://github.com/franciscovillaescusa/Quijote-simulations}, respectively. The analysis of the simulations and training of the neural network has been carried out in the Rusty cluster of the Flatiron Institute.


\appendix
\section{Appendix A: Method} \label{sec:appendixA}
In this section, we introduce a brief description of the convolutional neural networks and the \DM architecture used in this work.


\subsection{Convolutional neural networks}
\label{CNN}
An artificial neural network is an interconnected group of nodes (neurons) that combine to learn complex patterns. Neural networks are modeled after biological neural network systems and attempt to allow computers to learn in a similar manner to humans. An artificial neuron receives one or more input $x_i$ and multiplies them by a set of connection weights $w_i$ plus a bias. The result of such operation is called net input:
\begin{equation}
    z = \sum w_i x_i+b~.
\end{equation}

The output can be then calculated by applying the activation function over the net input. The activation function is a non-linear function which evaluates if the neuron would fire or not. There exist different types of activation functions, but the Rectified Linear Unit (ReLU) is the most widely used in neural networks nowadays. ReLU activation function is defined as:
\begin{equation*}
    f(x)=
     \begin{cases}
      0 &\text{if }  x < 0,\\
      x &\text{if } x \ge 0.
     \end{cases}
\end{equation*}
 One of the greatest advantage of ReLU is that it does not activate all neurons at the same time. It converts all negative inputs to zero and the neuron does not get activated. This makes it very computational efficient than other activation functions.
 
A particular class of neural networks that can be used to learn spatial features is called \textit{Convolutional Neural Networks} (CNN). A CNN consists of an input and an output layer, as well as multiple hidden layers. The input consists of images that computers read as pixels and express them as matrices. The convolutional layer contains a set of learnable parameters (also known as filters or kernels). A filter is used to detect the presence of specific features or patterns present in the input.
This filter is convolved across the input file, and a dot product is computed to give an activation map. The feature images produced by a convolutional layer is then passed to the next layer in the CNN. 

One of the main problems we can encounter when training learning systems is a change in the distribution of the outputs of each hidden layer. Each input layer is affected by the model parameters of all preceding layers. As a consequence, a small change made within a layer can produce a large variation of the inputs of another layer and change feature maps distribution. This means that later layers needs to continuously adapt to the new distribution obtained from the previous one. This change can slow down the convergence of the learning algorithm. In order to improve the performance of the neural network, we can make use of the Batch normalization technique~\cite{Ioffe}. Batch normalization (BN) consists in normalizing the output of a previous activation layer during the training, so that the input to the activation function across each training batch has a fixed mean (0) and a variance (1). Instead of using the entire data set to normalize activation functions, the normalization is restrained to each mini-batch in the training process.

\subsection{The \DM~Architecture}
We follow~\cite{He2019} and we use a \DM~architecture that consists of three parts: the contracting path, a bottleneck and the expansive path. During the contracting path, the spatial resolution is reduced while the feature information is increased. The expansive path increases the spatial resolution and combines it with the feature information. The bottleneck is in between the contracting and expansive path and allows the model to learn a compression of the input data.

The contracting and expansive paths consist each of blocks of two $3^3$ convolutional neural networks, while the bottom of the U-Net consists of two convolutional neural networks. Every convolution is followed by zero padding, BN, and ReLU. As in U-Net, at all resolution levels, with exception of the bottleneck, we concatenate the input of the downsampling layers to the output of the upsampling layers. This allows to transfer the small-scale spatial information from the contracting path to the expansive path, that it would otherwise have poor spatial resolution. At the final layer, a $1^3$ convolution is used to map 64 features to the final 3-D data. We refer the reader to~\cite{He2019} and~\cite{U-Net} for a more detailed description of the architecture. 

\section{Appendix B: Robustness of \DM~model}
\label{AppendixNN}

\begin{figure}[h]
\vspace{0.25cm}
\begin{center}
\includegraphics[width=0.5\textwidth]{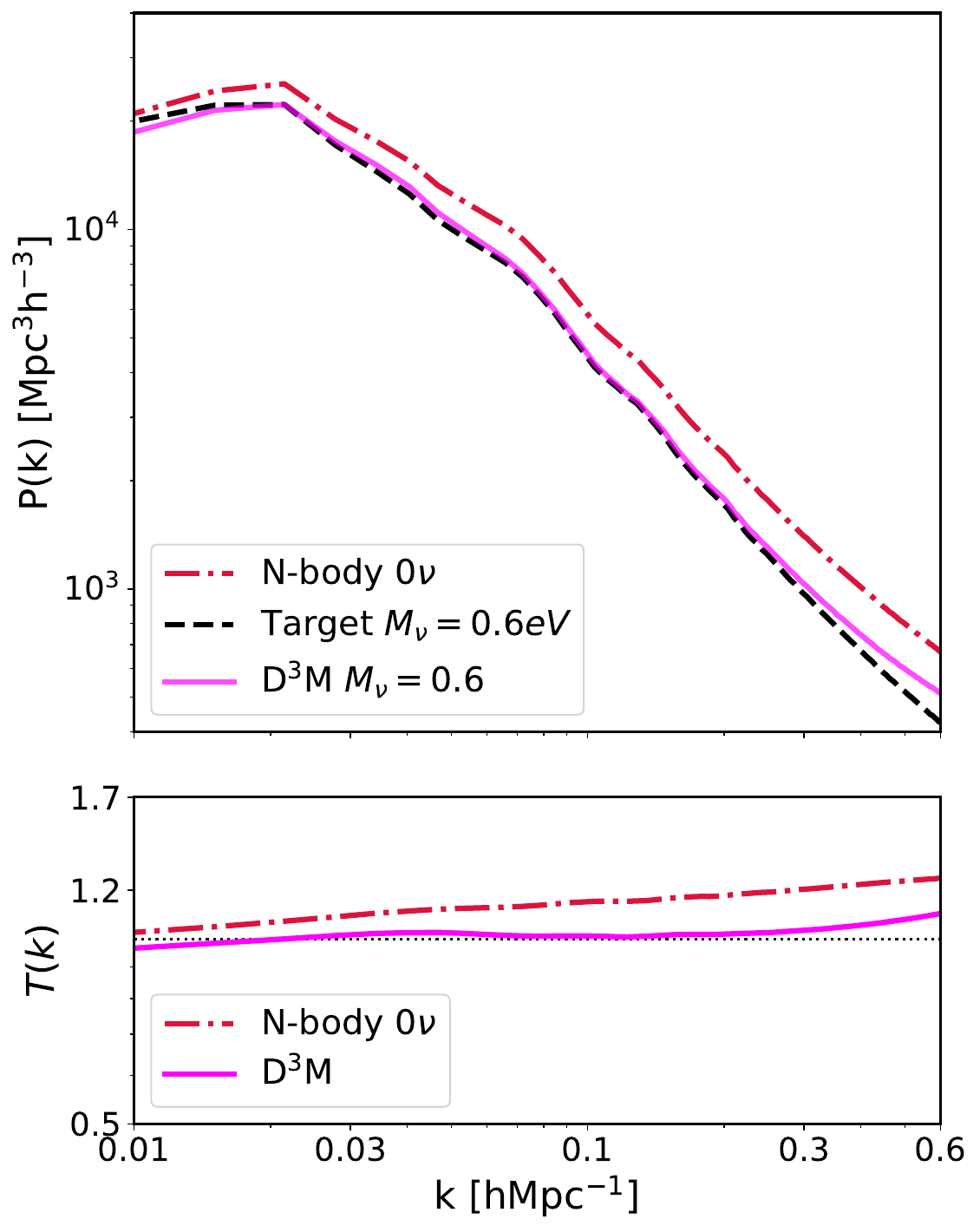}  
\caption{The top panel shows the matter power spectrum comparison among the standard N-body simulations (red dashed-dotted line), the new target (black dashed line) and the prediction (magenta line). The bottom panel depicts the transfer function for the input and the \DM.}
\label{fig:Pk2}
\end{center}
\end{figure}

In order to test the robustness of the \DM, we train our neural network by considering a different neutrino mass case. In particular, while the input still includes standard simulations without neutrinos, the new target consists of neutrino simulations with mass $M_\nu=0.6$~eV. The results are depicted in Figure~\ref{fig:Pk2}, where we show the power spectrum and the transfer function for the input, target and the prediction of the neural network.
Also for this neutrino mass case, we can note a very good agreement in the power spectrum between the target and the \DM~on large-scales and a deviation from the target on small non-linear scales. In particular, the transfer function is approximately 1 for $0.03 < k < 0.45$ h/Mpc and differs from the unity by a $8\%$ from scales $k\approx 0.5$~h/Mpc.
Those results suggest that our model is applicable independently of neutrino mass ensuring the robustness of our approach.\\

\bibliography{references}{}
\bibliographystyle{hapj}

\end{document}